\documentclass{iopart}
\usepackage{graphicx}
\usepackage{amssymb}
\usepackage{slashbox}
\def\ii{{\rm i}}
\def\nn{\nonumber}

\begin{document} 
\title[Unstable $g$-modes]{Unstable $g$-modes in Proto-Neutron Stars}

\author{V Ferrari$^1$, L Gualtieri$^1$ and J A Pons$^2$} 
\address{$^1$ Dipartimento di Fisica ``G.Marconi", 
Sapienza Universit\` a di Roma \\
and Sezione INFN  ROMA1, piazzale Aldo  Moro 2, I-00185 Roma, Italy} 
\address{$^2$ Departament de F\'{\i}sica Aplicada, Universitat d'Alacant,\\ 
Apartat de correus 99, 03080 Alacant, Spain}

\begin{abstract} 
In this article we study the possibility that, due to non-linear
couplings, unstable $g$-modes associated to convective motions excite
stable oscillating g-modes.  This problem is of particular interest,
since gravitational waves emitted by a newly born proto-neutron star
pulsating in its stable $g$-modes would be in the bandwidth of VIRGO
and LIGO.  Our results indicate that nonlinear saturation of unstable
modes occurs at relatively low amplitudes, and therefore, even if
there exists a coupling between stable and unstable modes, it does not
seem to be sufficiently effective to explain, alone, the excitation of
the oscillating $g$-modes found in hydrodynamical simulations.
\end{abstract} 
\section{Introduction} 
It is well known that proto-neutron stars (PNS) born in the aftermath
of a gravitational core collapse are, during the first tens of
seconds, convectively unstable \cite{Pons}. In addition, dynamical
simulations of core collapse have shown that a considerable fraction of
the energy emitted in gravitational waves (GW) is actually due to this
convective motion, rather than to the violent bounce
\cite{Muller,Ott04}.
Recently, a new mechanism for core collapse supernova explosion has
been proposed, based on the excitation and sonic damping of core
$g$-mode oscillations \cite{Burrows}-\cite{Bur07}, and the same
mechanism has also been pointed out to be relevant for GW emission
\cite{Ott06}.  The process begins being an advective-acoustic
oscillation, and in \cite{Bur06} the authors suggest that the primary
agent of the explosion is the acoustic power generated in the inner
turbulent region; the accreting PNS is a
self-excited oscillator, and  the core oscillation acts as a
transducer converting accretion energy into acoustic energy.  
Surprisingly, and
despite the complexity of the non-linear hydrodynamical effects
involved in these simulations, the growth times, oscillation
frequencies, and temporal evolution of the modes are qualitatively and
quantitatively similar to the predictions of a linear analysis done in
\cite{FMP}; there, the frequencies and damping times of $g$-modes were
computed for an evolving PNS, modeled as sequences
of quasi-stationary configurations.

These results put the $g$-modes in a somewhat different perspective
with respect to the preceeding literature on the subject; indeed,
previous studies of $g$--mode  oscillations in neutron stars had shown
that their contribution to the gravitational wave signal was likely
to be too weak to be detectable by current or near-future detectors
\cite{Finn87}-\cite{noi}.

Motivated by these considerations, in this paper we study the spectrum
of stable and unstable $g$-modes of a newly-born PNS, and we investigate
how effective the coupling is, due to the non linearity of the
equations of hydrodynamics, between unstable $g$-modes and oscillating
$g$-modes.  The aim is to understand whether this coupling  induces
$g$-mode oscillations which may significantly contribute to
gravitational wave emission during the early phases of a
proto-neutron star's life.  To this purpose, we integrate the
equations of relativistic stellar perturbations for the model of PNS
considered in \cite{FMP}, which describes the quasi-stationary evolution of
a PNS during the first minute after core collapse; we look for the
eigenmodes the frequency of which, at some point of the evolution,
becomes imaginary or acquires a negative imaginary part, a behaviour
which indicates a mode instability. We then discuss to what extent
unstable and stable $g$-modes are coupled.

It should be mentioned that the existence of an unstable branch of
$g$--modes is a natural consequence of the PNS thermal and chemical
profiles, since some regions of the star are convectively unstable while
other regions are not.  Thus, even at the linear level one can see
that both stable and unstable $g$-modes are present.  The linear
analysis allows to determine which is the fastest growing mode and the
frequencies of the oscillatory modes.

The  PNS we consider, at very early stage has a 
radius of about 100 km, and it gradually contracts in a few seconds.
It has a fixed baryonic mass of 1.6 $M_\odot$, which corresponds
to a final gravitational mass (after deleptonization and cooling)
of 1.46 $M_\odot$. The radius of the PNS at t=50 ms
after bounce is 50 km but just 1 second later, due to
neutrino losses from the envelope, the star contracts to
a radius of less than 20 km. The final radius of the cold,
catalyzed NS (after 30-40 seconds) is  12 km.
In computing the g-mode eigenfrequencies at selected values of time, 
we have taken the corresponding radial profiles 
(mass, pressure, temperature, lepton content, etc.)
provided by the evolutionary models of \cite{Pons},  already used in \cite{FMP}.

Evaluating the effect of nonlinear couplings between different modes
is a formidable task; however an estimate of the leading order
contributions can be provided using a Newtonian approach based on the
amplitude equation formalism developed in \cite{VH} and \cite{Goupil}.

The paper is organized as follows. In Section \ref{setup} we describe
the mathematical and numerical setup of our calculations; in Section
\ref{modes} we discuss the structure of stable and unstable $g$-modes
in PNS; in Section \ref{coupling} we study the coupling between
different modes in terms of appropriate integrals of the
radial eigenfunctions; concluding remarks are in Section \ref{concl}.

\section{Equations and numerical setup}\label{setup}
The equations governing the perturbations of spherical stars in general
relativity have been derived within different approaches by many
authors \cite{TC}-\cite{DL}. Here we use the Lindblom-Splinter (LS)
equations derived in \cite{LS}, which describe  dipolar ($l=1$)
perturbations, and the Lindblom-Detweiler (LD) equations derived in
\cite{LD}, for $l\ge 2$ perturbations.
The metric and the fluid four-velocity of the spherical background
describing the unperturbed star are
\begin{eqnarray}
ds^2&=&-e^\nu dt^2+e^\lambda dr^2+r^2(d\theta^2+\sin^2\theta d\phi^2)\nn\\
u^\mu&=&(e^{-\nu/2},0,0,0)\,,\label{background}
\end{eqnarray}
where the functions $\nu(r)$, $\lambda(r)$ satisfy the TOV equations
(see for instance \cite{MTW}). We shall assume that matter in the star
is a perfect fluid.

The metric perturbations and the Eulerian fluid perturbations are
expanded in spherical harmonics $Y^{lm}(\theta,\phi)$, and
Fourier-expanded as described in \cite{LD,LS}. In particular,
the radial component of the Lagrangian displacement is
\begin{equation}
\delta r=\delta r^{lm}(\omega,r)Y^{lm}(\theta,\phi) e^{\ii\omega t}\,.
\end{equation}
The perturbed Einstein equations
\[
\delta G_{\mu\nu}=8\pi \delta T_{\mu\nu}\label{einstein}
\]
provide a system of linear, ordinary differential equations (ODE) for
the radial components of the harmonic expansion of the metric and fluid
perturbations. Since the structure of the equations for $l=1$ and for $l>1$
is quite different, we will treat the two cases separately.

$l=1$  perturbations are not associated to
gravitational wave emission. Therefore, since we are neglecting
further dissipative effects like viscosity or heat transport,
oscillation modes are undamped, {\it normal modes} and the
corresponding eigenfrequencies, $\omega_k$, are real.
Dipolar oscillations are described by the LS equations, which form a
linear, third order ODE system \cite{LS}.  In order to find the
mode eigenfrequencies, these equations have to be integrated from the
center (where regularity conditions must be imposed) up to the stellar
surface, for different values of $\omega$.  The solutions satisfy
the required boundary condition at the surface, i.e. the vanishing of
the Lagrangian perturbation of the pressure, only for a discrete set of
values of $\omega$: these are the normal mode eigenfrequencies.  The
equations depend on $\omega$ quadratically, and since all terms in
these equations and all boundary conditions are {\it real}, there
is no restriction in assuming that the perturbation functions and
$\omega^2$ are real, i.e. in looking for normal modes.  Unstable modes
correspond to solutions belonging to $\omega^2<0$. They are non
oscillating, exponentially growing modes, with growth time $\tau$
given by
\begin{equation}
\frac{1}{\tau}=\sqrt{-\omega^2}\,.
\end{equation}
For $l\ge2$, stellar
perturbations are associated to gravitational wave emission, therefore
the corresponding modes, said {\it quasi-normal modes}, are damped and
the eigenfrequencies are complex
\begin{equation}
\omega=2\pi\nu+\frac{\ii}{\tau}\,;
\end{equation}
$\nu$ is the mode frequency, and $\tau$ is its gravitational damping
time.  These perturbations are described by the LD equations, which
form a linear, fourth order  ODE system \cite{LD}.  These equations
are integrated from the center, imposing regularity conditions, to the
stellar surface, for different values of complex $\omega$. For each
$\omega$ there exist two regular, independent solutions which have to
be matched on the stellar surface to allow the vanishing of the
Lagrangian perturbation of the pressure.  This constraint reduces the number of
independent solutions to one, modulo an overall scaling factor.

Outside the star, the variables associated to the fluid motion vanish
and the equations reduce to a second-order ODE for the Zerilli
function $Z^{lm}(\omega,r_*)$ \cite{Zer}
\begin{equation}
\label{zer} \left(\frac{d^2}{dr_*^2} +\omega^2\right) Z^{lm}(\omega,r_*) =
V^l(r)Z^{lm}(\omega,r_*)~,
\end{equation}
where $V^l(r)$ is the Zerilli potential and 
$r_*=r +2M \log\left(\frac{r}{2M}-1 \right)$ is the 
tortoise coordinate.  Far away from the star, the Zerilli function
behaves as a superposition of ingoing and ingoing waves:
\begin{equation}
Z^{lm}(r)\simeq Z^{lm}_{in}e^{\ii\omega r_*}+
Z^{lm}_{out}e^{-\ii\omega r_*}\,.\label{infinity}
\end{equation}
Quasi-normal modes are the solutions of the LD+Zerilli equations for which the
ingoing component of the asymptotic solution (\ref{infinity}),
$Z_{in}^{lm}$, vanishes. The discrete set of frequencies corresponding
to these solutions are the mode eigenfrequencies.

Like the LS equations, the LD equations have real coefficients;
however the outgoing wave boundary condition $Z_{in}^{lm}=0$ cannot be
satisfied by a real $Z^{lm}$. Therefore, in order to represent a
physical mode the harmonic components of the perturbations must belong
to complex frequencies.  For $l\ge2$ unstable modes are solution of
the perturbed equations with $\Im(\omega)<0$; an unstable mode corresponds to a
perturbation which, while oscillating, is increasingly growing in
amplitude; the oscillation frequency $\nu$ and the growth time $\tau (> 0)$
of such unstable mode are given by $\omega=2\pi\nu-{\ii}/{\tau}$.

The relevant equations for $l\ge1$ have been integrated using an
adaptive stepsize fourh-order Runge-Kutta method. For $l\ge 2$ the
solution has been continued outside the star by integrating 
Zerilli's equation (\ref{zer}) using the continued fraction method
\cite{LNS};  the complex eigenfrequencies have been found as zeroes
of the function $Z^{lm}_{in}(\omega)$, using a Newton-Raphson method.

\section{Normal and quasi-normal modes of proto-neutron stars}\label{modes}
Before discussing our results, we shall briefly summarize what is
known in the literature about the mode structure of non rotating stars
in Newtonian gravity.  This will serve as a guideline to understand
our results for relativistic PNS.

The normal mode spectrum of non-rotating stars has been studied in
great detail in Newtonian gravity (see for instance \cite{Cox},\cite{otherNewt}). The
main results can be summarized as follows.
\begin{itemize}
\item Since Newtonian stars do not emit gravitational waves, in
absence of viscous forces the square of the mode  eigenfrequencies, $\nu_k^2$, are
real for any value of $l$.  The index $k$ denotes the order of the
mode.  If $\nu_k^2>0$, the mode corresponds to a stable oscillation,
while if $\nu_k^2<0$ the mode is a pure exponential growth.
\item Mode eigenfrequencies group in distinct classes, identified on
the basis of the restoring force which dominates in bringing back to
the equilibrium position the generic fluid element displaced by the
perturbation; they are named $g_k$-modes if the restoring force is
buoyancy, $p_k$-modes if it is due to pressure gradients.  $g$-mode
eigenfrequencies are smaller than those of $p$-modes, and the two
classes are separated by the single fundamental mode ($f$-mode)
frequency.
\item When $l=1$, the $f$-mode frequency is identically zero: the
$f$-mode does exist only for $l\ge2$.  $l=1$ $g$- and $p$-modes have an interesting
characteristic:  they correspond to a displacement of the geometrical center of
the system, i.e. $\delta r(0)\neq 0$; as pointed out in
\cite{Smeyers}, in the $l=1$ case the geometrical center does not
coincide with the center of mass of the star, which is not displaced.
\item There can exist two classes of $g$-modes: the $g_k^{(+)}$ modes,
which are stable, and the $g_k^{(-)}$, which are unstable. Whether
both branches of modes do exist or not depends on the sign of the
Schwarzschild discriminant
\begin{equation}
A=\frac{\rho'}{\rho}-\left(\frac{d\rho}{dp}\right)_s
\frac{p'}{\rho}\,.\label{SD}
\end{equation}
\begin{itemize}
\item If $A=0$ throughout the star (as it is for barotropic EOS), then
there are no $g$-modes (or, more precisely, they are all degenerate to
zero frequency).
\item If $A<0$ throughout the star (convective stability), there are
only $g^{(+)}$-modes.
\item If $A>0$ throughout the  star (convective instability), 
there are only $g^{(-)}$-modes.
\item If $A<0$ in some region of the star, and $A>0$ in some other
region, then both $g^{(+)}$-modes and $g^{(-)}$-modes are present.
\end{itemize}
Thus, the sign of the Schwarzschild discriminant reveals the presence
of convective instabilities, and of unstable $g$-modes.
$p$-modes and the $f$-mode are always stable.
\item Typically, 
for $g^{(\pm)}_k$- and $p_k$-modes
the number of nodes in the eigenfunction of $(\delta r)_k$ 
is equal to $k$ (not counting the node at $r=0$ for $l\neq1$). 
The $f$-mode eigenfunction has no nodes inside the star. 
However, such simple prescriptions may not hold in some cases \cite{Cox}.
\end{itemize}

Let us now consider our relativistic PNS.  In order to find the
frequencies of the quasi-normal modes, we have integrated the
equations that describe the perturbations of a spherical star taking
as a background the ``quasi--stationary" evolutionary models derived
in \cite{Pons}.  ``Quasi--stationary'' means that the stellar interior
is described by a sequence of equilibrium configurations, which have
been shown to adequately describe the evolution of a PNS for $t
\gtrsim 0.1-0.2$ s after bounce \cite{Pons,pons2}.  These
models have been used in \cite{FMP} to compute how the
quasi-normal mode frequencies of stable, quadrupole modes evolve in
time. Here we consider the equation of state named GM3, with no quark
matter in the core, and explore the time interval from $t=0.2$s up to
$t=25$s after bounce.  We find  both stable and unstable
$g$-modes.  Unstable modes are present because the relativistic
Schwarzschild discriminant \cite{ThorneIII}
\begin{equation}
S(r) = \frac{dp}{dr} - \left(\frac{\partial p}{\partial\epsilon}
\right)_s\frac{d\epsilon}{dr}
\label{RSD}
\end{equation}
is negative in some region inside the star, revealing the presence of
convective instability\footnote{Notice that the sign convention for
the relativistic discriminant (\ref{RSD}) is the opposite to that 
assumed for the Newtonian discriminant (\ref{SD}).}.

\begin{figure}[ht]
\begin{center}
\includegraphics[width=6cm,angle=270]{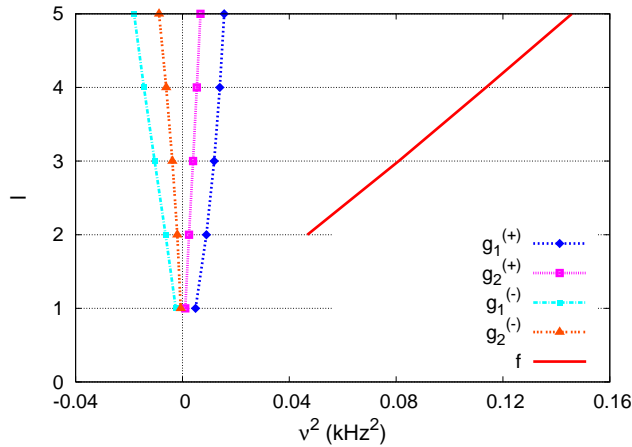}
\caption{Mode structure of a PNS at $t=2s$ after bounce.}
\label{FIG1}
\end{center}
\end{figure}

To compare our results with the literature on Newtonian stars (in
particular, with Figure 1 of \cite{Cox}), we define a real, square
frequency, $\nu^2$, as follows.
\begin{itemize}
\item For $l=1$, $\omega_k^2$ is always real; it is positive for
stable modes, negative for unstable modes. Therefore we define
$\nu_k^2\equiv(\omega_k/2\pi)^2$.
\item For $l\ge2$, $\omega_k$ is complex.  For stable fluid modes, in
general $|\Re{\omega_k}|\gg|\Im{\omega_k}|$, whereas for unstable
modes $|\Im{\omega_k}|\gg|\Re{\omega_k}|$.  Therefore, we can define
\begin{eqnarray}
(2 \pi \nu_k)^2&\equiv&(\Re\omega_k)^2\simeq\omega_k^2
~~~~~\hbox{for stable modes}\nn\\
(2 \pi \nu_k)^2&\equiv&-(\Im\omega_k)^2\simeq\omega_k^2
~~~~~\hbox{for unstable modes}\,.
\end{eqnarray}
\end{itemize}
The mode structure of the considered stellar model is shown in Figure
\ref{FIG1}, where $\nu_k^2$ is plotted for different values of the
harmonic index $l$.
$\nu_k^2$ is plotted for the first two  unstable $g$-modes,
$g^{(-)}_1$ and $g^{(-)}_2$, respectively, for the first two  stable $g$-modes,
$g^{(+)}_1$ and $g^{(+)}_2$, and for the fundamental mode.
The figure refers to $t=2$ s after bounce.\\
\noindent
We see that $\nu^2$ has a dependence on $l$
similar to that shown in the Newtonian case \cite{Cox}.  We remind that the
main difference is that for $l>1$ in the relativistic case
$\omega_k^2$ is complex: the frequency of stable modes acquires a
small imaginary part corresponding to the damping induced by
gravitational wave emission, while that of unstable modes acquires a
small real part.

\begin{figure}[ht]
\begin{center}
\includegraphics[width=6cm,angle=270]{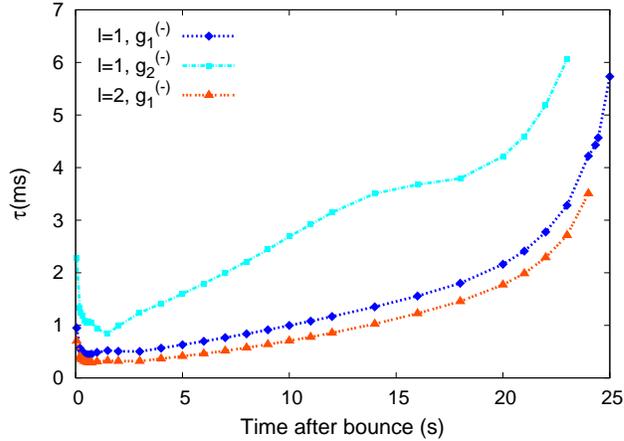}
\caption{Growth time of the lowest unstable $g$-modes
during the first 25 seconds of the PNS life. \label{ttau}}
\label{FIG2}
\end{center}
\end{figure}

Let us now focus on the behaviour of unstable $g$-modes. The lowest dipole
mode, (i.e. $g^{(-)}_1$ for $l=1$),  at $t=2$ s after bounce has a growth time
$\tau_{g_1^{(-)}}=0.51$ ms. This is very interesting because, since
$\tau_{g_1^{(-)}}$ is extremely small, the amplitude of this
mode may have enough time to grow significantly before
being damped by other dissipative effects, or by non-linear couplings.
For $l\ge 2$, the growth times of the lowest unstable $g$-modes
are even smaller; for instance, for $l=2$, $\tau_{g_1^{(-)}}=0.3$ ms.
In Figure \ref{FIG2} we show how the growth time of the lowest
unstable $g$-modes for $l=1,2$ changes during the first $25$ s of the PNS
life. We see that, after an initial decrease,
all $\tau_{g_i^{(-)}}$'s increase with time; however, they remain quite small,
being their maximum values  of the order of a few ms.
After about  $25$ s,  the star becomes convectively stable and unstable
modes disappear. 

For $l\ge 2$ unstable $g$-modes  have
a small oscillating part, which is of the order of $\nu\sim
10^{-8}$ Hz and can be neglected.

\section{Mode couplings}\label{coupling}
The existence of rapidly growing, unstable $g$-modes indicates, at a
linear level, the onset of an effective convective instability; this motivates
our further investigation on mode coupling, 
to see whether unstable modes, coupling to stable ones, may trigger
gravitational wave emission at frequencies which fall in the bandwidth of ground based
interferometers, as shown in \cite{FMP}.

\begin{figure}[ht]
\begin{center}
\includegraphics[width=6cm,angle=270]{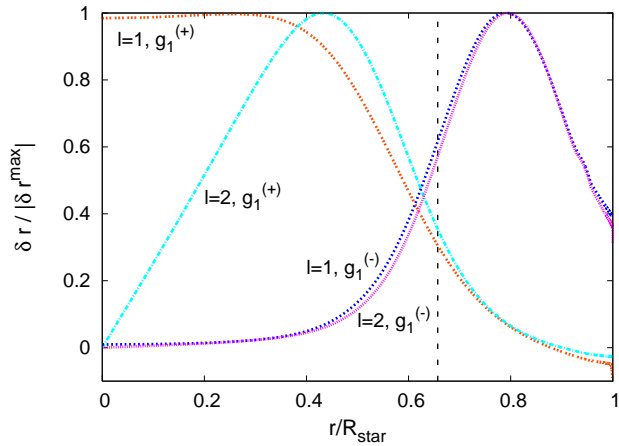}
\caption{Eigenfunctions of the lowest stable and unstable $g$-modes, plotted for
$l=1,2$ as  functions of the normalized radius.  The Schwarzschild discriminant $S(r)$ is
positive on the left of the dashed vertical line, negative on the
right. This picture refers to $t=0.5$s after bounce.}
\label{FIG3}
\end{center}
\end{figure}

A first thing one can do is to
see whether the corresponding eigenfunctions overlap: if they do not,
we would have a strong indication that the coupling is ineffective.
In Figure \ref{FIG3} we compare, as an example, the radial
eigenfunctions of the lowest, stable and unstable $g$-modes belonging to
$l=1,2$.
We see that there is indeed an overlap\footnote{Notice that the
eigenfunctions of the stable modes are defined mostly where $S(r)>0$,
while the eigenfunctions of the unstable modes are defined mostly
where $S(r)<0$.}; however, this information is not sufficient to infer
how effective the transfer of energy from convective motion to
oscillatory modes can be.

In order to quantify the effectiveness of the couplings, we have used
the amplitude equations formalism described in \cite{VH} and
\cite{Goupil}, where  couplings between different modes have  been analysed in the
framework of Newtonian gravity.
In this approach, the Lagrangian displacement $\vec{\delta r}(t,\vec r)$ is
expressed as a superposition of modes
\begin{equation}
\vec{\delta r}(t,\vec r)=\sum_ia^i(t)\vec{\xi}_{(i)}(\vec r) ,
\end{equation}
where $\vec{\xi}_{(i)}(\vec r)$ are the mode eigenfunctions, normalized
by imposing $\xi_{(i)}^r(R_{star})/R_{star}=1$, and $a^i(t)$ are the
corresponding amplitudes.  For non-resonant modes
\footnote{Resonant modes are those for which the condition $\sum_j
~k_j \nu_j \simeq 0$ is satisfied, with $k_j$ small integers
\cite{Goupil}.}, the non-linear coupling between different modes is
described by the equation
\begin{equation}
\dot a^i=\ii\omega_i a^i+Q^i_{iii}(a^i)^3+Q^i_{ijj}a^i(a^j)^2 ,
\label{aeq}
\end{equation}
where the frequency $\omega_i$ is in general complex,
i.e. $\omega_i=2\pi \nu_i\pm\frac{\ii}{\tau_i}$, if some dissipative
process is in action.  In our case, dissipation is due to
gravitational waves and, as mentioned before, for stable modes we can
assume $2 \pi \nu \gg\frac{1}{\tau}$ while for unstable modes $2 \pi
\nu \ll\frac{1}{\tau}$.

The evaluation of the coupling coefficients $Q^i_{ijj}$, the
expressions of which are given in \cite{VH} and \cite{Goupil}, is a very hard
task.  Since we need an order of magnitude estimate, we shall compute
only a leading term.  In particular, we shall use the expressions of
$Q^i_{ijj}$ given in \cite{Goupil}, but we have checked that using
those given in \cite{VH} the results are of the same order of
magnitude.  The leading term we consider is
\begin{equation}
Q^i_{ijj}=\ii C^i_{ijj}\frac{3(2-\delta_{ij})Z^i_{ijj}}{
2\omega_iI^i_i}+\dots
\label{qii}
\end{equation}
where
\begin{eqnarray}
Z^i_{ijj}&=&\int d\Omega Y^*_iY_iY_jY_j\nn\\
I^i_i&=&\int dr\rho r^2\left[|\xi^r_{(i)}|^2+l(l+1)|\xi^\perp_{(i)}|^2
\right]\nn\\
C^i_{ijj}&=&\int dr\rho r^2\frac{p'}{\rho}(\xi^r_{(i)})^*
\left(\frac{\delta \rho_{(i)}}{\rho}\right)
\left(\frac{\delta \rho_{(j)}}{\rho}\right)^2\,.\label{exprC}
\end{eqnarray}
$Y_i$ is the spherical harmonic corresponding to the $i$-th mode, and
$\rho, p, \delta\rho, \delta p$ are the density, the pressure, and
their Lagrangian perturbations.  If all quantities are expressed in
geometrical units ($G=c=1$) as powers of km, then $Q^i_{ijj}$ has
dimension km$^{-1}$, which can be converted to $s^{-1}$ multiplying
by $c$.

Let us first compute the coefficients $Q^i_{iii}$, i.e. the
self-coupling coefficients (see eq. (\ref{aeq})), for an unstable
mode.  Since in this case $\omega_i\simeq-\frac{i}{\tau_i}$,
eq. (\ref{qii}) gives
\begin{equation}
Q^i_{iii}=-\frac{3\tau}{2}Z^i_{iii}\frac{C^i_{iii}}{I^i_i}+\dots ~,
\end{equation}
and, for the lowest unstable $g$-modes corresponding to $l=1,2$, we find
\begin{eqnarray*}
&&|Q^{l=1}| \sim 3.0\cdot 10^{13}~ {\rm s}^{-1}\\
&&|Q^{l=2}| \sim 1.3\cdot 10^7~ {\rm s}^{-1} .
\end{eqnarray*}
Assuming that $Q^i_{iii}<0$, i.e. that the effect of self-coupling is
that of saturating the mode, using the values of $|Q^{l=1}|$ and
$|Q^{l=2}|$ found above, we can estimate the saturation amplitude as
the value of $a^i$ for which $\dot a^i=0$, i.e.
\begin{equation}
a^i_{sat}\simeq\frac{1}{\sqrt{|Q^i_{iii}|\tau}}\,.
\end{equation}
We find:
\begin{eqnarray*}
&\hbox{for}\quad l=1~~ &\delta r(R_{star}) \sim 8.2~ {\rm cm}\\
&\hbox{for}\quad l=2~~ &\delta r(R_{star}) \sim 150~ {\rm m}~.
\end{eqnarray*}
\begin{table}[ht]
\begin{center}
\caption{Dimensionless coupling coefficients $\tilde Q^i_{ijj}(a^i_{sat})$
between stable modes (index $i$) $g^{(+)}_{nl}$, $f_l$ and unstable
modes (index $j$) $g^{(-)}_{nl}$.\label{tabQ}}
\begin{tabular}{|c|c|c|c|}\hline
\backslashbox{$i$}{$j$} & $g^{(+)}_{12}$ & $g^{(+)}_{22}$ &
$f_{2}$ \\\hline
$g^{(-)}_{11}$ & $4\cdot 10^{-5}$& $2\cdot 10^{-7}$& $6\cdot10^{-5}$\\
\hline
$g^{(-)}_{12}$ & $ 10^{-2}$& $2\cdot 10^{-6}$& $2\cdot10^{-2}$\\
\hline
$g^{(-)}_{21}$ & $4\cdot 10^{-5}$& $2\cdot 10^{-8}$& $6\cdot10^{-5}$\\
\hline
$g^{(-)}_{22}$ & $2\cdot 10^{-2}$& $6\cdot 10^{-7}$& $4\cdot10^{-2}$\\
\hline
\end{tabular}
\end{center}
\end{table}
Next we want to check whether unstable modes can trigger stable modes,
eventually making them unstable.  Therefore, we need to evaluate the
coefficients which describe the coupling between the $i$-th stable
mode and the $j$-th unstable mode.  If the $i$-th mode is stable,
eq. (\ref{aeq}) becomes
\begin{equation}
\dot a^i\simeq \ii 2 \pi \nu_i a^i-\frac{1}{\tau_i}a^i
+Q^i_{iii}(a^i)^3+Q^i_{ijj}a^i(a^j)^2\,.\label{coupleq}
\end{equation}
The purely imaginary terms on the right-hand side of (\ref{coupleq})
determine a shift of the frequency $\nu_i$ and are not relevant for
the mode stability; therefore, we are interested only in the real part
of the right-hand side of (\ref{coupleq}).  If we define
\begin{equation}
\tilde Q^i_{ijj}\equiv
C^i_{ijj}\frac{3Z^i_{ijj}}{(2 \pi \nu_i)^2I^i_i}\,,
\end{equation}
then eq. (\ref{aeq}) can be written as
\begin{equation}
\dot a^i\simeq\ii\left(\dots\right)-\frac{1}{\tau_i}a^i\left[
1+\tilde Q^i_{ijj}(a^j)^2\right]\,.
\label{adot}
\end{equation}
We must then evaluate the dimensionless quantity $\tilde
Q^i_{ijj}(a^j_{sat})^2$, where $a^j_{sat}$ is the saturation amplitude
of the unstable mode $j$, previously computed.  In particular, we want
to check whether the sign of the term in square brackets in equation
(\ref{adot}) can be reversed.

It should be noted that the stable $i$-th mode must belong to $l\ge
2$, because modes belonging to $l=1$ are not associated to
gravitational wave emission, and for them $1/\tau_i=0$. This means
that the mechanism of mode enhancing we are studying is triggered by
gravitational wave emission, as in the case of the well-known CFS
instability \cite{CFS}.

In Table \ref{tabQ} we show the values of $|\tilde
Q^i_{ijj}|~(a^j)^2$; the index $j$ refers to unstable modes
$g^{(-)}_{nl}$, where $n=1,2$ is the order of the mode, and $l=1,2$;
the index $i$ refers to stable modes $g^{(+)}_{nl}$ and $f_l$
(fundamental mode), with $l=2$.  In all cases we find
\begin{equation}
|\tilde Q^i_{ijj}|(a^j_{sat})^2\ll 1 ;
\end{equation}
this means that the term in square brackets cannot change sign and
that the effect of unstable modes on stable modes is too weak to
induce a significant growth of their oscillation amplitudes.

\section{Concluding remarks}\label{concl}
We have determined the frequencies of stable and unstable modes of a
newly born proto-neutron star during the early phases of its life.
Our goal was to understand whether the coupling between stable and
unstable modes could provide a mechanism to convert the energy of
convective motion into $g$-mode oscillations, enhancing gravitational
wave emission.

Our results indicate that the existing coupling do not seem large
enough to explain, alone, the excitation of the $g$-mode oscillations
found in hydrodynamical simulations.  Thus, other processes should be
considered as driving the excitation of these modes, as, for instance,
accretion from the mantle \cite{Burrows}, \cite{Yoshida}.

\ack We are grateful to Tim Van Hoolst for suggesting us how to
estimate the order of magnitude of the coupling coefficients.  We
thank Harald Dimmelmeier and Andrea Passamonti for useful discussions.

 
\end{document}